\documentclass[lettersize,journal]{IEEEtran}

\IEEEoverridecommandlockouts

\usepackage{textcomp}
\usepackage{stfloats}
\usepackage{verbatim}
\usepackage{cite}
\usepackage{amsmath,amssymb,amsfonts}
\usepackage{algorithmic}
\usepackage{graphicx}
\usepackage{float}
\usepackage{textcomp}
\usepackage{xcolor}
\usepackage{bbm}
\usepackage{array}
\usepackage{rotating}
\usepackage{gensymb}
\usepackage{comment}
\usepackage[acronym]{glossaries}
\usepackage{multirow}
\usepackage{lscape}
\usepackage{makecell}
\usepackage[inline]{enumitem}
\usepackage{svg}
\usepackage{hyperref}
\hypersetup{hidelinks}
\usepackage[font=small,labelsep=period,justification=centering]{caption}
\usepackage{subcaption}

\usepackage{fancyhdr}

\usepackage{adjustbox}
\usepackage{tikz}
\usetikzlibrary{shapes,arrows}
\usetikzlibrary{positioning}
\usetikzlibrary{decorations.text}

\newcounter{CorrCounter}
\newcounter{LemmaCounter}
\newcounter{PropCounter}

\DeclareMathOperator*{\argmin}{arg\,min}

\def\BibTeX{{\rm B\kern-.05em{\sc i\kern-.025em b}\kern-.08em
    T\kern-.1667em\lower.7ex\hbox{E}\kern-.125emX}}

\allowdisplaybreaks


\begin{document}

\onecolumn

\title{Electromagnetic Modelling of Extended Targets in a Distributed Antenna System}

\author{\IEEEauthorblockN{Baptiste Sambon, François De Saint Moulin, Guillaume Thiran, Claude Oestges, and Luc Vandendorpe\thanks{François De Saint Moulin, Baptiste Sambon, and Guillaume Thiran are Research Fellows of the Fonds de la Recherche Scientifique - FNRS.}}\\
\IEEEauthorblockA{ICTEAM, UCLouvain - Louvain-la-Neuve, Belgium\\}
email: firstname.lastname@uclouvain.be
}

\maketitle

\thispagestyle{plain}%
\pagestyle{plain}

\begin{abstract}
Traditional radar and integrated sensing and communication (ISAC) systems often approximate targets as point sources, a simplification that fails to capture the essential scattering characteristics for many applications. This paper presents a novel electromagnetic (EM)-based framework to accurately model the near-field (NF) scattering response of extended targets, which is then applied to three canonical shapes : a flat rectangular plate, a sphere and a cylinder. Mathematical expressions for the received signal are provided in each case. %
Based on this model, the influence of bandwidth, carrier frequency and target distance on localisation accuracy is analysed, showing how higher bandwidths and carrier frequencies improve resolution. Additionally, the impact of target curvature on localisation performance is studied. Results indicate that detection performance is slightly enhanced when considering curved objects. 
A comparative analysis between the extended and point target models shows significant similarities when targets are small and curved. However, as the target size increases or becomes flatter, the point target model introduces estimation errors owing to model mismatch. 
The impact of this model mismatch as a function of system parameters is analysed, and the operational zones where the point abstraction remains valid and where it breaks down are identified.
These findings provide theoretical support for experimental results based on point-target models in previous studies.

\end{abstract}


\section{Introduction}
\label{sec:introduction}

Sixth-Generation (6G) systems are expected to integrate Localisation and Sensing (L\&S) with communications to form Integrated Sensing and Communications (ISAC) systems \cite{chen_6g_2024}. In this context, sensing performance can be improved by larger bandwidths at higher carrier frequencies and the use of Extremely Large Antenna Arrays (ELAAs) \cite{ye_extremely_2024}.  
As the frequency and antenna array dimensions increase, the Fraunhofer distance, i.e. the Far Field (FF) limit proportional to the square of the array size and the carrier frequency, grows, expanding the Near Field (NF) region. 

Most work in radar and ISAC abstracts the target as a single point, which is not valid in the NF region as the target appears spatially extended \cite{liu_crb_2024, wang_cramer-rao_2024, cong_near-field_2024}. As an example, experimental results in \cite{khosravi_experimental_2024} highlight the impact of multiple backscattering points at mmWave frequencies, emphasising the need for models that accurately capture extended target responses in the NF. 
Unfortunately, to the best of the authors' knowledge, a quantitative characterisation of the conditions under which a point target model is sufficient or an extended target model becomes necessary is still lacking.

In the NF region, the spherical nature of the wavefront must be taken into account, adding complexity due to the varying channel gain across antennas. However, this NF effect may offer new possibilities, i.e. by allowing spatial localisation rather than just angular estimation \cite{liu_near-field_2024}. By exploiting NF propagation, both numerical and experimental studies \cite{sakhnini_near-field_2022, sakhnini_experimental_2022} have demonstrated improved localisation accuracy. However, the target models used in these studies are restricted to point targets, which limits their ability to capture the full scattering characteristics of extended objects and may lead to biased estimates due to model mismatch.

In \cite{moulin_near-field_2024}, a new NF electromagnetic (EM)-based model is proposed for a flat rectangular plate, which is then used to derive a Maximum Likelihood (ML) estimator for range. The model is based on the Stationary Phase Approximation (SPA) \cite{bleistein1975asymptotic} and provides a closed form expression for the received signal. By using both amplitude and phase information, the ML estimator shows improved performance compared to traditional range estimators. In addition, performance bounds for this model are derived in \cite{thiran_performance_2024}, showing a degradation in performance when the flat plate is approximated as a single point. %
However, the scope of the studies presented in these articles is limited: only a flat rectangular plate is considered as the target, with the antenna array constrained to be centred and oriented parallel to the plate surface.

\subsection{Contributions}

Based on the aforementioned works, the contributions of this paper are summarised as follows.
\begin{itemize}
    \item The NF EM-based model of \cite{moulin_near-field_2024} is first generalised to allow for arbitrary target geometries and antenna positions in a multi-static scenario, providing a general closed form expression for the received signal in the case of extended targets. The model is particularised to three canonical shapes: a flat rectangular plate, a spherical target, and a cylindrical target.
    \item Based on this generalised model, numerical analyses are performed for the estimation of the distance between an antenna array and the target, for the three considered target geometries. Namely, they show that similar resolutions are obtained for curved and flat geometries, but lower sidelobe levels are observed in the ambiguity function for curved targets.
    \item A stationary point analysis is also performed, highlighting variations in the density of stationary points based on the curvature of the object. This analysis provides insight into the limitations of the point target approximation. 
    \item A novel range estimator is developed to perform a model mismatch analysis where the target, though extended, is approximated as a single point. In particular, this analysis derives the range bias introduced by the point target approximation as a function of system parameters, enabling the identification of operational zones where the point target approximation remains valid and where it fails. This underscores the necessity for an extended target model in those cases.
\end{itemize}

\subsection{Structure of the paper}

The paper is structured as follows. First, the general NF EM-based model is presented in Section~\ref{sec:system_model}. Then, the Stationary Phase Approximation (SPA) used to obtain closed-form expressions is introduced in Section~\ref{sec:SPA}, followed by its application to a flat rectangular target in Section~\ref{sec:flat_rectangular_target}, and to curved targets in Section~\ref{sec:curved_target}. In Section~\ref{sec:numerical_analysis}, numerical analyses are performed for the estimation of the distance for the three canonical targets and the impact of curvature is analysed. The impact of the model mismatch on range estimation and a stationary point analysis are presented in Section~\ref{sec:stationary_points_analysis}, while a detailed characterization of the mismatch as a function of system parameters is provided in Section~\ref{sec:model_mismatch_characterisation}.

\section{Electromagnetic-based Sensing}
\label{sec:system_model}
\begin{figure}
    \centering
    \includegraphics[width=0.4\textwidth]{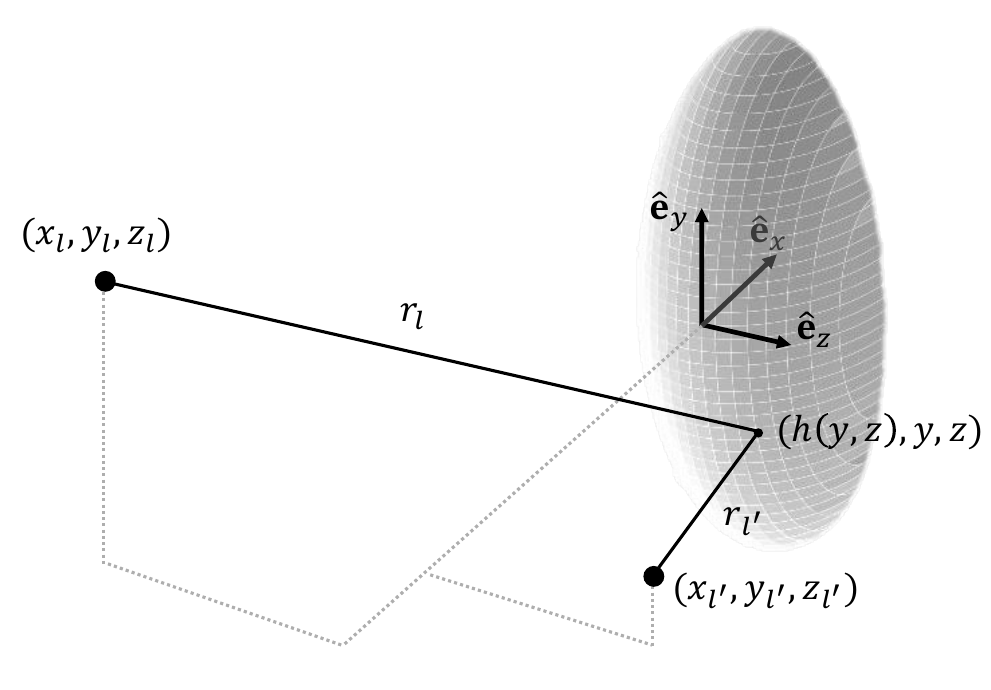}
    \caption{Reflection on a target between two antennas.}
    \label{fig:system_model}
\end{figure}

The multi-static scenario illustrated in Figure \ref{fig:system_model} is considered. The target $\mathcal{S}$ reflects the signals transmitted by a set of $N$ antennas defined as 
\begin{equation}
\mathcal{A} \triangleq \left\{(x_l,y_l,z_l) \in \mathbb{R}^3,\: l = 0,...,N-1 \right\}.
\end{equation}
The shape of the target is modelled through an arbitrary function $h : \mathbb{R} \times \mathbb{R} \rightarrow \mathbb{R}$, such that
\begin{equation}
\mathcal{S} \triangleq \left\{(x,y,z) \in \mathbb{R}^3 \: \left| \: 
    x = h(y,z) \right. \right\}.
\end{equation}
It is also assumed that the function $h$ varies slowly w.r.t. the wavelength. Thus, the distance $r_l$ from the $l^\text{th}$ antenna to any point $(x,y,z) \in \mathcal{S}$ is computed as 
\begin{equation}
r_l = \sqrt{(h(y,z)-x_l)^2 + (y-y_l)^2 + (z-z_l)^2}. \label{eq:distance}
\end{equation}
Moreover, the normal unit vector pointing outside the target at the point $(x,y,z) \in \mathcal{S}$ is given by
\begin{equation}
\hat{\textbf{e}}_n = \frac{-\hat{\textbf{e}}_x + \frac{\partial h}{\partial y} \: \hat{\textbf{e}}_y + \frac{\partial h}{\partial z} \: \hat{\textbf{e}}_z}{\sqrt{1 + \left(\frac{\partial h}{\partial y}\right)^2 + \left(\frac{\partial h}{\partial z}\right)^2}}.
\end{equation}
One can also define the spherical coordinate system associated with the $l^\text{th}$ antenna as
\begin{equation}
\left[\begin{matrix}
    \hat{\textbf{e}}_{r_l} \\
    \hat{\textbf{e}}_{\theta_l} \\
    \hat{\textbf{e}}_{\phi_l}
\end{matrix}\right] \triangleq \left[\begin{matrix}
    \cos \theta_l \cos \phi_l & \sin \theta_l & \cos \theta_l \sin \phi_l \\
    -\sin \theta_l \cos \phi_l & \cos \theta_l & -\sin \theta_l \sin \phi_l \\
    -\sin \phi_l & 0 & \cos \phi_l
\end{matrix}\right] \left[
    \begin{matrix}
        \hat{\textbf{e}}_x \\
        \hat{\textbf{e}}_y \\
        \hat{\textbf{e}}_z
    \end{matrix}\right],
\end{equation}
where $\theta_l$ and $\phi_l$ are the elevation and azimuth angles. The elevation angle is computed from the horizontal plane, and the azimuth angle from the $x$-axis. These angles can be computed for any point $(x,y,z) \in \mathcal{S}$ from the following equalities:
\begin{equation}
    \begin{alignedat}{10}
        &\sin \phi_l &&= \frac{z-z_l}{\sqrt{r_l^2 - (y-y_l)^2}}, \:\: &&\sin \theta_l &&= \frac{y-y_l}{r_l}, \\
        &\cos \phi_l &&= \frac{h(y,z)-x_l}{\sqrt{r_l^2 - (y-y_l)^2}}, \:\: &&\cos \theta_l &&= \frac{\sqrt{r_l^2 - (y-y_l)^2}}{r_l}.
    \end{alignedat}
\end{equation}

Assuming that the antennas can be modeled as linear antennas axed along $\hat{\textbf{e}}_y$, the incident fields from the $l^\text{th}$ antenna at the point $(x,y,z) \in \mathcal{S}$ of the target are computed as \cite{orfanidis_electromagnetic_2016}
\begin{equation}
\begin{alignedat}{10}
&\textbf{E}^i_l(t) &&= - j k \eta L I_0 \: s_l\left(t-\frac{r_l}{c}\right) \: f(\theta_l) \: \frac{e^{-jk r_l}}{4\pi r_l} \: \hat{\textbf{e}}_{\theta_l}, \\
&\textbf{H}^i_l(t) &&= \frac{1}{\eta} \: \hat{\textbf{e}}_{r_l} \times \textbf{E}_l^i(t),
\end{alignedat}
\label{eq:incident_fields}
\end{equation}
with $k = 2\pi/\lambda$ the wave number, $\lambda$ the wavelength, $\eta$ the free-space impedance, $L$ the antenna length, $I_0$ the current in the antenna, $s_l$ the signal transmitted by the $l^\text{th}$ antenna, $c$ the speed of light, and $f(\theta_l)$ the normalised radiation pattern of the antenna. The transmit signal is assumed to have a bandwidth $B << c/\lambda$. Equation \eqref{eq:incident_fields} is valid for $r_l >> \lambda$, which is usually always the case at the considered carrier frequencies, e.g. for automotive radar systems.\\

Following the physical optics approximation \cite{gutierrez-meana_high_2011}, assuming that the target is a good electrical conductor (which is well approximated by a perfect electrical conductor), the equivalent surface current density generated by the $l^\text{th}$ antenna at the surface of the target is computed from the incident magnetic field as 
\begin{align}
\nonumber \textbf{J}_l(t) &= 2 \: \hat{\textbf{e}}_n \times \textbf{H}_l^i(t) \\
&= -2 j k L I_0 \: s_l\left(t-\frac{r_l}{c}\right) \: f(\theta_l) \: \: \frac{e^{-jk r_l}}{4\pi r_l} \: \left(\hat{\textbf{e}}_n \times \hat{\textbf{e}}_{\phi_l}\right).
\end{align}
The noise-free received signal at the ${l'}^\text{th}$ antenna from the ${l}^\text{th}$ antenna is then computed from the equivalent surface current density as \cite{orfanidis_electromagnetic_2016}
\begin{equation}
u_{l'l}(t) = -jk\eta \iint_\mathcal{S} \left(\textbf{J}_{l'l}^\perp\left(t-\frac{r_{l'}}{c}\right)  \cdot \textbf{f}^\perp_{l'}\right) \: \frac{e^{-jkr_{l'}}}{4\pi r_{l'}} \: \text{d} \mathcal{S}, \label{eq:received_signal_full_1}
\end{equation}
with $\textbf{J}_{l'l}^\perp(t) \triangleq \textbf{J}_l(t) - \left(\textbf{J}_l(t) \cdot \hat{\textbf{e}}_{r_{l'}}\right) \hat{\textbf{e}}_{r_{l'}}$ the component of the equivalent surface current density perpendicular to the direction of propagation, and $\textbf{f}^\perp_{l'} \triangleq L f(\theta_{l'}) \: \hat{\textbf{e}}_{\theta_{l'}}$ the antenna effective length. In order to simplify \eqref{eq:received_signal_full_1}, one may notice that $\hat{\textbf{e}}_{r_{l'}} \cdot \hat{\textbf{e}}_{\theta_{l'}} = 0$, and that $\left(\hat{\textbf{e}}_n \times \hat{\textbf{e}}_{\phi_l}\right) \cdot \hat{\textbf{e}}_{\theta_{l'}} = \hat{\textbf{e}}_n \cdot \left(\hat{\textbf{e}}_{\phi_{l}} \times \hat{\textbf{e}}_{\theta_{l'}}\right)$, leading to
\begin{equation}
    u_{l'l}(t) = -\frac{k^2\eta L^2 I_0}{8\pi^2} \iint_\mathcal{S} g_{l'l}(y,z) \: e^{j \psi_{l'l}(y,z)} \: \text{d} \mathcal{S}, \label{eq:received_signal_full}
\end{equation}
where the amplitude function $g_{l'l}$ and the phase function $\psi_{l'l}$ are respectively defined as 
\begin{align}
g_{l'l}(y,z) &\triangleq s_{l}\left(t- \frac{r_l + r_{l'}}{c}\right) \: \frac{f(\theta_l)f(\theta_{l'})}{r_l r_{l'}} \: \left[\hat{\textbf{e}}_n \cdot \left(\hat{\textbf{e}}_{\phi_{l}} \times \hat{\textbf{e}}_{\theta_{l'}}\right)\right], \\
\psi_{l'l}(y,z) &\triangleq - k (r_l + r_{l'}). \label{eq:phase_function}
\end{align}

\section{Stationary Phase Approximation}
\label{sec:SPA}
The Stationary Phase Approximation (SPA) is a powerful tool for approximating integrals of oscillatory functions. For univariate integrations, the SPA method is described in \cite{moulin_near-field_2024}. The extension to multivariate integrations is presented here.

\subsection{Multivariate Integration}   
The SPA can also be extended to multivariate integrals of the form
\begin{equation}
    I = \int_\Omega \textbf{g}(\textbf{x}) \: e^{j\psi(\textbf{x})} \: \text{d}\textbf{x},
\end{equation}
where $\textbf{x} \in \Omega \subseteq \mathbb{R}^n$. The functions $\textbf{g}$ and $\psi$ are respectively named amplitude and phase functions. The stationary point $\textbf{x}_s$ is still obtained by nullifying the gradient of the phase function, i.e.
\begin{equation}
\left.\nabla \psi(\textbf{x})\right|_{\textbf{x}=\textbf{x}_s} = 0. \label{eq:stationary_point_definition}
\end{equation}
Then, assuming that the amplitude function varies slowly in the vicinity of the stationary point, and that the integration domain is large enough to contain multiple oscillations of the phase function, following the same development as in the univariate case, the integral is approximated by
\begin{equation}
    I \approx \textbf{g}(\textbf{x}_s) \: e^{j\psi(\textbf{x}_s)} \: (2\pi)^{\frac{n}{2}} \: \left|\det\left\{\nabla^2\psi(\textbf{x}_s)\right\}\right|^{-\frac{1}{2}} \: e^{j\frac{\pi}{4}\text{sgn}\left\{\nabla^2\psi(\textbf{x}_s)\right\}},
\label{eq:spa_multi_expression}
\end{equation}
where $\text{det}\left\{\cdot\right\}$ and $\text{sgn}\left\{\cdot\right\}$ are respectively the determinant and the signature of the Hessian matrix of the phase function. In that case, since the Hessian matrix is symmetric, the signature is either $n$ if the stationary point is a minimum, or $0$ if it is a saddle point. %
It should be noted that, if there is multiple stationary points, the integral is again approximated by the sum of the contributions of the amplitude function around each stationary point.\\

In this paper, we assume that both assumptions required to apply the SPA are fulfilled. Yet, if the integration domain is not large enough to contain multiple oscillations of the phase function, the last approximation does not hold. In that case, the integration of the phase function may be performed using the Fresnel function \cite{abramowitz_handbook_1965}.

\subsection{Computation of the stationary point}
The stationary point, denoted as $\textbf{x}_s = (x_s,y_s,z_s)$, may be located by solving the system of equations provided by \eqref{eq:stationary_point_definition}. Based on the phase function defined as \eqref{eq:phase_function}, the stationary point should satisfy the following system of equations: 
\begin{equation}
\left\{
    \begin{alignedat}{10}
        &\left.\frac{\partial \psi_{l'l}}{\partial y}\right|_{\substack{y=y_s \\ z= z_s}} = 0, \\
        &\left.\frac{\partial \psi_{l'l}}{\partial z}\right|_{\substack{y=y_s \\ z= z_s}} = 0.
    \end{alignedat}
\right. \label{eq:stationary_system_undeveloped}
\end{equation}
Let us drop the indices $l'l$ of the phase function and specular points associated with each pair of antennas for the sake of readability. Denoting 
\begin{alignat}{10}
    &h_s &&\equiv h(y_s,z_s) = &x_s,&&  \\ 
    &h'_{ys} &&\equiv \left.\frac{\text{d}h}{\text{d}y}\right|_{\substack{y=y_s \\ z= z_s}}, \: &h'_{zs} &&\equiv \left.\frac{\text{d}h}{\text{d}z}\right|_{\substack{y=y_s \\ z= z_s}}, 
\end{alignat}
and 
\begin{equation}
    r_{ls} \equiv \sqrt{(h_s-x_l)^2 + (y_s-y_l)^2 + (z_s-z_l)^2}, \label{eq:distance}
\end{equation}
the different quantities evaluated at the stationary point, one can show that \eqref{eq:stationary_system_undeveloped} is developed as \eqref{eq:stationary_system_developed}.\\

\begin{figure*}
\begin{equation}
    \left\{
        \begin{alignedat}{10}
            &\frac{(h_s - x_l)\:h_{ys}'  + (y_s-y_l)}{r_{ls}} + \frac{(h_s - x_{l'})\:h_{ys}' + (y_s - y_{l'})}{r_{l's}} = 0, \\
            &\frac{(h_s - x_l)\:h_{zs}' + (z_s-z_l)}{r_{ls}} + \frac{(h_s - x_{l'})\:h_{zs}' + (z_s - z_{l'})}{r_{l's}} = 0.
        \end{alignedat}
    \right. \label{eq:stationary_system_developed}
\end{equation} 
\hrule
\end{figure*}

Note that different coordinate systems may be considered for each pair of antennas, and chosen such that $h_s = 0$, and/or $h_{ys}' = h_{zs}' = 0$. In that case, this system of equations can be further simplified. Nonetheless, for many target and antennas geometries, it is difficult to solve these equations analytically. Fortunately, following the Fermat stationary point theorem \cite{miller_fermats_2009}, the stationary point minimises the total travelled distance $r_l + r_{l'}$. Additionally, the stationary point can also be computed knowing that it corresponds to the specular reflection point, i.e. the point for which the angle of incidence and reflection on the plane are equal.

\subsection{Computation of the Hessian Matrix}
Based on the phase function defined as \eqref{eq:phase_function}, the Hessian matrix can be computed from the first and second derivatives of the function $h$. For the sake of readability, the indices $l'l$ of the phase function and the specular points associated with each pair of antennas are omitted again. Denoting 
\begin{equation}
    h''_{ys} \equiv \left.\frac{\text{d}^2h}{\text{d}y^2}\right|_{\substack{y=y_s \\ z= z_s}}, \: h''_{zs} \equiv \left.\frac{\text{d}^2h}{\text{d}z^2}\right|_{\substack{y=y_s \\ z= z_s}},\: h''_{yzs} \equiv \left.\frac{\text{d}^2h}{\text{d}y\text{d}z}\right|_{\substack{y=y_s \\ z= z_s}},
\end{equation}
the different quantities evaluated at the stationary point, one can show that the Hessian matrix of the phase function is computed as 
\begin{equation}
    \nabla^2\psi(\textbf{x}_s) = \left[
        \begin{matrix}
            \left.\frac{\partial^2 \psi}{\partial y^2}\right|_{\textbf{x}=\textbf{x}_s} & \left.\frac{\partial^2 \psi}{\partial y\partial z}\right|_{\textbf{x}=\textbf{x}_s} \\
            \left.\frac{\partial^2 \psi}{\partial y\partial z}\right|_{\textbf{x}=\textbf{x}_s} & \left.\frac{\partial^2 \psi}{\partial z^2}\right|_{\textbf{x}=\textbf{x}_s}
        \end{matrix}
    \right],
\end{equation}
where the different elements of the matrix are given by \eqref{eq:d2psidy2}, \eqref{eq:d2psidz2}, and \eqref{eq:d2psidydz}. Again, note that different coordinate systems may be considered for each pair of antennas, and chosen such that $h_s = 0$, and/or $h_{ys}' = h_{zs}' = 0$. In that case, these expressions can be further simplified.

\begin{figure*}
\begin{align}
\nonumber    \left.\frac{\partial^2 \psi}{\partial y^2}\right|_{\textbf{x}=\textbf{x}_s} &= \frac{k}{r_{ls}^3} \left[y_s-y_l + (h_s - x_l)  \: h_{ys}'\right]^2 - \frac{k}{r_{ls}} \left[1+h_{ys}'^2 + (h_s - x_l) \: h_{ys}''\right] \\
&+ \frac{k}{r_{l's}^3} \left[y_s-y_{l'} + (h_s - x_{l'})  \: h_{ys}'\right]^2 - \frac{k}{r_{l's}} \left[1+h_{ys}'^2 + (h_s - x_{l'}) \: h_{ys}''\right] \label{eq:d2psidy2}\\ 
\nonumber    \left.\frac{\partial^2 \psi}{\partial z^2}\right|_{\textbf{x}=\textbf{x}_s} &= \frac{k}{r_{ls}^3} \left[z_s-z_l + (h_s - x_l)  \: h_{zs}'\right]^2 - \frac{k}{r_{ls}} \left[1+h_{zs}'^2 + (h_s - x_l) \: h_{zs}''\right] \\
&+ \frac{k}{r_{l's}^3} \left[z_s-z_{l'} + (h_s - x_{l'})  \: h_{zs}'\right]^2 - \frac{k}{r_{l's}} \left[1+h_{zs}'^2 + (h_s - x_{l'}) \: h_{zs}''\right] \label{eq:d2psidz2}\\
\nonumber    \left.\frac{\partial^2 \psi}{\partial y\partial z}\right|_{\textbf{x}=\textbf{x}_s} &= \frac{k}{r_{ls}^3} \left[y_s-y_l + (h_s - x_l)  \: h_{ys}'\right]\left[z_s-z_l + (h_s - x_l)  \: h_{zs}'\right]- \frac{k}{r_{ls}} \left[h_{ys}' h_{zs}' + (h_s - x_l) \: h''_{yzs}\right] \\
&+ \frac{k}{r_{l's}^3} \left[y_s-y_{l'} + (h_s - x_{l'})  \: h_{ys}'\right]\left[z_s-z_{l'} + (h_s - x_{l'})  \: h_{zs}'\right]- \frac{k}{r_{l's}} \left[h_{ys}' h_{zs}' + (h_s - x_{l'}) \: h''_{yzs}\right] \label{eq:d2psidydz}
\end{align}
\hrule
\end{figure*}

\section{Application to a Flat Rectangular Target}
\label{sec:flat_rectangular_target}
Let us first consider a flat rectangular plate. While \cite{moulin_near-field_2024} focuses on an antenna array centred and aligned parallel to the plate, this work extends the analysis to arbitrary antenna positions relative to the plate. Without any loss of generality, the shape function $h$ is defined as $h(y,z) = 0$, such that 
\begin{equation}
\mathcal{S} \triangleq \left\{(x,y,z) \in \mathbb{R}^3 \: \left|\:
\begin{alignedat}{10}
    x = 0 \\
    |y| \leq \frac{D_y}{2}, \: |z| &&\leq \frac{D_z}{2}
\end{alignedat}
\right.\right\},
\end{equation}
with $D_y$ and $D_z$ the dimension of the plate along the $y-$ and $z-$axis, respectively. The indices $l'l$ of the phase functions and the specular points associated with each pair of antennas are still dropped here. Although the stationary point may, in general, lie outside the plate, we assume in this work that it is located on the plate, i.e., $\textbf{x}s \in \mathcal{S}$. Additionally, we assume that all antennas are positioned on the same side of the plate, such that $x_{l} < 0$ or $x_{l} > 0 \:\forall\: l = 0, \ldots, N-1$. While this assumption is not strictly necessary in practice, it significantly simplifies the following analytical calculations.

\subsection{Computation of the Stationary Point}
With this target model, since $\textbf{x}_s \in \mathcal{S}$ and $h(x,y) = 0 \: \forall \: (x,y) \in \mathcal{S}$, it follows that $h_s = 0$, $h_{ys}' = h_{zs}' = 0$. The system of equations given in \eqref{eq:stationary_system_developed} is thus simplified as 
\begin{equation}
    \left\{
        \begin{alignedat}{10}
            &\frac{y_s-y_l}{r_{ls}} + \frac{y_s - y_{l'}}{r_{l's}} = 0, \\
            &\frac{z_s-z_l}{r_{ls}} + \frac{z_s - z_{l'}}{r_{l's}} = 0.
        \end{alignedat}
    \right. 
\end{equation} 
Unfortunately, except for specific antenna configurations, the solution of this system of equations is not straightforward. Yet, knowing that the stationary point is also the specular reflection point, for the considered target and antennas positions, the following property should hold: 
\begin{equation}
\frac{x_{l}}{r_{ls}} = \frac{x_{l'}}{r_{l's}} \quad\Leftrightarrow\quad \frac{x_l}{x_{l'}} = \frac{r_{ls}}{r_{l's}},
\label{eq:x_l_x_l_prime}
\end{equation}
leading to 
\begin{equation}
y_s = \frac{y_l x_{l'} + y_{l'}x_l}{x_l + x_{l'}}, \: z_s = \frac{z_l x_{l'} + z_{l'}x_l}{x_l + x_{l'}}. \label{eq:specular_point_plate}
\end{equation} 
One can notice that, if a planar antenna array parallel to the plate is considered, the stationary point is obtained as the arithmetic mean of the antennas positions on the plane.

\subsection{Computation of the Hessian Matrix}
With this target model, since $\textbf{x}_s \in \mathcal{S}$ and $h(x,y) = 0 \: \forall \: (x,y) \in \mathcal{S}$, it also follows that $h_{ys}'' = h_{zs}'' = h_{yzs}'' = 0$. Thus, using \eqref{eq:specular_point_plate}, \eqref{eq:distance} is rewritten as 
\begin{equation}
r_{ls} = \frac{2x_l}{x_l+x_{l'}}\underbrace{\frac{1}{2}\sqrt{(x_l+x_{l'})^2 + (y_l - y_{l'})^2 + (z_l - z_{l'})^2}}_{\equiv \: r_{l'l}}.
\end{equation}
Consequently, \eqref{eq:d2psidy2}, \eqref{eq:d2psidz2}, and \eqref{eq:d2psidydz} are simplified as 
\begin{align}
    \left.\frac{\partial^2 \psi}{\partial y^2}\right|_{\textbf{x}=\textbf{x}_s} &= -\frac{k}{r_{l'l}^3} \frac{(x_l + x_{l'})^2}{8 x_l x_{l'}} \left[(x_l + x_{l'})^2 + (z_l - z_{l'})^2\right], \\
    \left.\frac{\partial^2 \psi}{\partial z^2}\right|_{\textbf{x}=\textbf{x}_s} &= -\frac{k}{r_{l'l}^3} \frac{(x_l + x_{l'})^2}{8x_l x_{l'}} \left[(x_l + x_{l'})^2 + (y_l - y_{l'})^2\right], \\
    \left.\frac{\partial^2 \psi}{\partial y\partial z}\right|_{\textbf{x}=\textbf{x}_s} &= \frac{k}{r_{l'l}^3} \frac{(x_l + x_{l'})^2}{8x_l x_{l'}} (y_l - y_{l'})(z_l-z_{l'}).
\end{align}
This is further simplified when a linear antenna array is considered along the $z$- or $y$-axis. For instance, in the first case, $y_{l} = 0 \: \forall \: l=0,...,N-1$, leading to
\begin{equation}
r_{l'l} = \frac{1}{2}\sqrt{(x_l+x_{l'})^2 + (z_l - z_{l'})^2},
\end{equation}
and
\begin{align}
    \left.\frac{\partial^2 \psi}{\partial y^2}\right|_{\textbf{x}=\textbf{x}_s} &= -\frac{k}{r_{l'l}} \frac{(x_l + x_{l'})^2}{2 x_l x_{l'}}, \\
    \left.\frac{\partial^2 \psi}{\partial z^2}\right|_{\textbf{x}=\textbf{x}_s} &= -\frac{k}{r_{l'l}^3} \frac{(x_l + x_{l'})^4}{8 x_l x_{l'}}, \\
    \left.\frac{\partial^2 \psi}{\partial y\partial z}\right|_{\textbf{x}=\textbf{x}_s} &= 0.
\end{align}
Consequently, the determinant is also easily computed as 
\begin{equation}
\det\left\{\nabla^2 \psi(\textbf{x}_s)\right\} = \frac{k^2}{r_{l'l}^4} \frac{(x_l + x_{l'})^6}{16 x_l^2 x_{l'}^2}.
\end{equation}
Applying the SPA, this leads back to the results obtained in \cite{moulin_near-field_2024}, considering that $x_l = x_{l'} = -R$ in this paper.

\section{Application to Curved Targets}
\label{sec:curved_target}

In the previous section it was shown that the stationary phase approximation can be used to derive an analytical expression for the scattered field of a flat rectangular target. In this section, this analysis is extended to curved targets, specifically considering cylindrical and circular targets. It is found that the SPA method can also be used to derive analytical expressions for the scattered fields of both curved target types.

\subsection{Application to a Spherical Target}

For a sphere, the shape function is defined as
\begin{equation}
    h(y,z) = \rho- \sqrt{\rho^2 - y^2 - z^2},
\end{equation}
where $\rho$ is the radius of the sphere. The surface of the sphere is then defined as
\begin{equation}
\mathcal{S} \triangleq \left\{(x,y,z) \in \mathbb{R}^3 \: \left|\:
\begin{aligned}
    x = &\rho- \sqrt{\rho^2 - y^2 - z^2} \\
    &|y| \leq \rho,  |z| \leq \rho
\end{aligned}
\right.\right\}. 
\end{equation}
The coordinate system is defined such that $h(0,0)=0$. 

The gradient of the shape function is given by
\begin{equation}
    \nabla h(y,z) = \begin{bmatrix}
        \frac{y}{\sqrt{\rho^2 - y^2 - z^2}} \\
        \frac{z}{\sqrt{\rho^2 - y^2 - z^2}}
    \end{bmatrix}.
\end{equation}
 
Similarly to the flat rectangular target, assuming the stationary point lies on the surface of the sphere, i.e. $\textbf{x}_s \in \mathcal{S}$, the stationary points are determined by solving \eqref{eq:stationary_system_developed}. For a spherical target, this system is simplified to
\begin{equation}
\left\{ 
    \begin{aligned}
        & \frac{(\rho- x_l) y_s - y_l d_{s}}{r_{ls}} + \frac{(\rho- x_{l'}) y_s - y_{l'} d_{s}}{r_{l's}} = 0, \\
        & \frac{(\rho- x_l) z_s - z_l d_{s}}{r_{ls}} + \frac{(\rho- x_{l'}) z_s - z_{l'} d_{s}}{r_{l's}} = 0,
    \end{aligned}
\right. 
\end{equation}
where $d_{s} = \sqrt{\rho^2 - y_s^2 - z_s^2}$. Since the analytical solution to these equations is difficult to compute, the coordinates of the stationary points can be computed numerically by minimising the total distance travelled $r_l + r_{l'}$ leveraging Fermat's Theorem \cite{miller_fermats_2009}.

The Hessian matrix of the shape function $h$ is computed as 
\begin{equation}
    \nabla^2 h(y,z) = \begin{bmatrix}
        \frac{\rho^2 - y^2}{(\rho^2 - y^2 - z^2)^{3/2}} & \frac{yz}{(\rho^2 - y^2 - z^2)^{3/2}} \\
        \frac{yz}{(\rho^2 - y^2 - z^2)^{3/2}} & \frac{\rho^2 - z^2}{(\rho^2 - y^2 - z^2)^{3/2}}
    \end{bmatrix}.
\end{equation}
Based on these expressions, the SPA defined in \eqref{eq:spa_multi_expression} can be used to calculate the signal $u_{l'l}$ at the receiving antenna by summing the contributions of the two stationary points. 

\subsection{Application to a Cylindrical Target}

In the case of a cylindrical target, the surface can be described by 
\begin{equation}
    \mathcal{S} \triangleq \left\{(x,y,z) \in \mathbb{R}^3 \: \left|\:
        \begin{aligned}
            x &= \rho - \sqrt{\rho^2 - z^2} \\
            &|y| \leq \frac{L}{2},  |z| \leq \rho
        \end{aligned}
\right.\right\}, 
\end{equation}
where $L$ and $\rho$ are the length and radius of the cylinder, respectively. The coordinate system is chosen so that $h(0,0)=0$ and the origin is at mid-length. 

The lack of curvature in the $y$ direction results in zero derivatives along that axis, resulting in $h'_{ys} = h''_{ys} = h''_{yzs} = 0$. Meanwhile, the derivative in the $z$-direction is the same as that derived for the spherical target. Consequently, the stationary points can be calculated as follows:
\begin{equation}
    \left\{
    \begin{aligned}
        &\frac{y_s-y_l}{r_{ls}} + \frac{y_s - y_{l'}}{r_{l's}} = 0, \\
        & \frac{(\rho- x_l) z_s - z_l d_{c}}{r_{ls}} + \frac{(\rho- x_{l'}) z_s - z_{l'} d_{c}}{r_{l's}} = 0,
    \end{aligned}
    \right. 
\end{equation}
where $d_c$ is here equal to $d_c=\sqrt{\rho^2 - z_s^2}$. The coordinates of the stationary points can be calculated numerically by using Fermat theorem of stationary points \cite{miller_fermats_2009}. 
To compute the determinant of the Hessian matrix of the phase function $\psi(y,z)$ evaluated at the stationary point, the following expression for the second derivatives in the $z$-direction must be used: 
\begin{equation}
    h_{zs}^{''} = \frac{\rho^2}{(\rho-z_s^2)^{3/2}}. 
\end{equation}

The requisite expressions for the SPA problem can then be employed to evaluate the integral \eqref{eq:received_signal_full}, using  \eqref{eq:spa_multi_expression}.

\section{Impact of Parameters}
\label{sec:numerical_analysis}

This section presents a numerical validation and analysis of the EM model.  
The signals \( s_{l} \) are assigned to orthogonal resources and have a waveform defined by \( s_{l}(t) = \frac{\sin(\pi B t)}{\pi B t} \), where \( B \) represents the signal bandwidth. Note that, in all generality, those signals can be communication signals in case of ISAC applications. 
The antennas are arranged linearly along an axis parallel to the $z$-axis and positioned at $y = 0$. The positions of the antennas are given by \( x_l = - R \) and \( z_l = \left(-\frac{N - 1}{2} + l \right) \Delta \), where $\Delta$ is the antenna spacing. From now, it is assumed that the elevation and azimuth angles from the centre of the array to the target are both equal to zero. Unless stated otherwise, the scenario parameters are those listed in Table \ref{tab:parameters}.

\begin{table}
    \caption{Scenario Parameters}
    \centering
    \begin{tabular}{|c|c|}
        \hline 
        Number of antennas & $N=13$ \\
        \hline 
        Antenna spacing & $\Delta=0.125 \, \text{m}$ \\
        \hline 
        Antenna length/current & $L^2 I_0=1 \, \text{m}^2 \, \text{A}$ \\
        \hline 
        Signal bandwidth & $B=100 \, \text{MHz}$ \\
        \hline 
        Carrier frequency & $f_c=77 \, \text{GHz}$ \\
        \hline 
        Target dimensions & \begin{tabular}{l}
        $D_y=0.8 \, \text{m}$ \\
        $D_z=1.75 \, \text{m}$ \\
        $\rho = 1.24 \, \text{m}$
        \end{tabular} \\
        \hline 
        Target position & $R=4\, \text{m}$ \\
        \hline
        Elevation angle & $\theta=0 \, \text{°}$ \\
        \hline
        Azimuth angle & $\phi=0 \, \text{°}$ \\
        \hline
    \end{tabular}
    \label{tab:parameters}
\end{table}

The impact of the bandwidth $B$, carrier frequency $f_c$ and distance $R$ from the target on the log-ML functions of the range estimator introduced in \cite{moulin_near-field_2024} for the flat rectangular plate is here studied for the cylindrical and spherical targets. The range is here defined as the closest distance from the centre of the antenna array to the target surface. The profile function of the estimation problem is analysed, i.e. the value of the log-ML function is evaluated at range $\tilde{R}$ for a target located at range ${R}$, in a noise-free scenario.

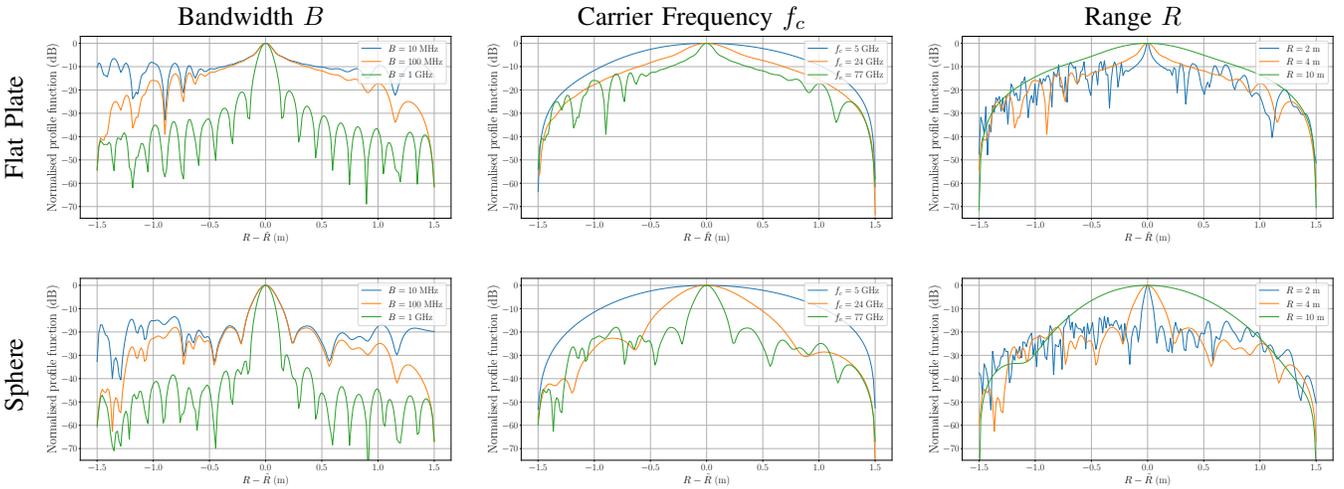
\begin{figure*}[ht]
    \centering
    \begin{tabular}{ccc}
        Bandwidth $B$ & Carrier Frequency $f_c$ & Range $R$ \\
        \begin{subfigure}[t]{0.3\textwidth}
            \centering
            \includegraphics[width=\linewidth, trim=60 0 100 60, clip]{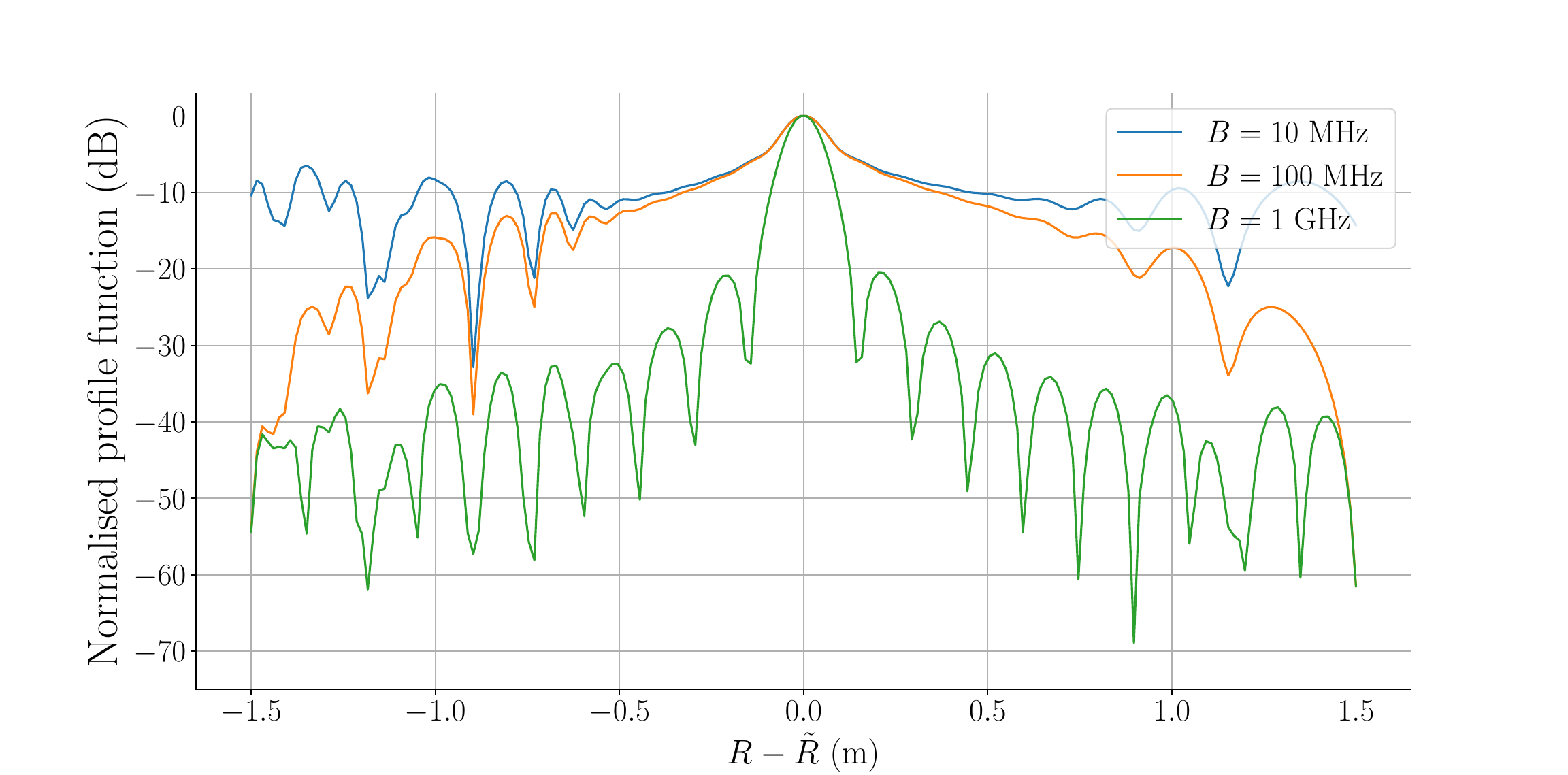}
            \vspace{0.01pt}
        \end{subfigure} &
        \begin{subfigure}[t]{0.3\textwidth}
            \centering
            \includegraphics[width=\linewidth, trim=60 0 100 60, clip]{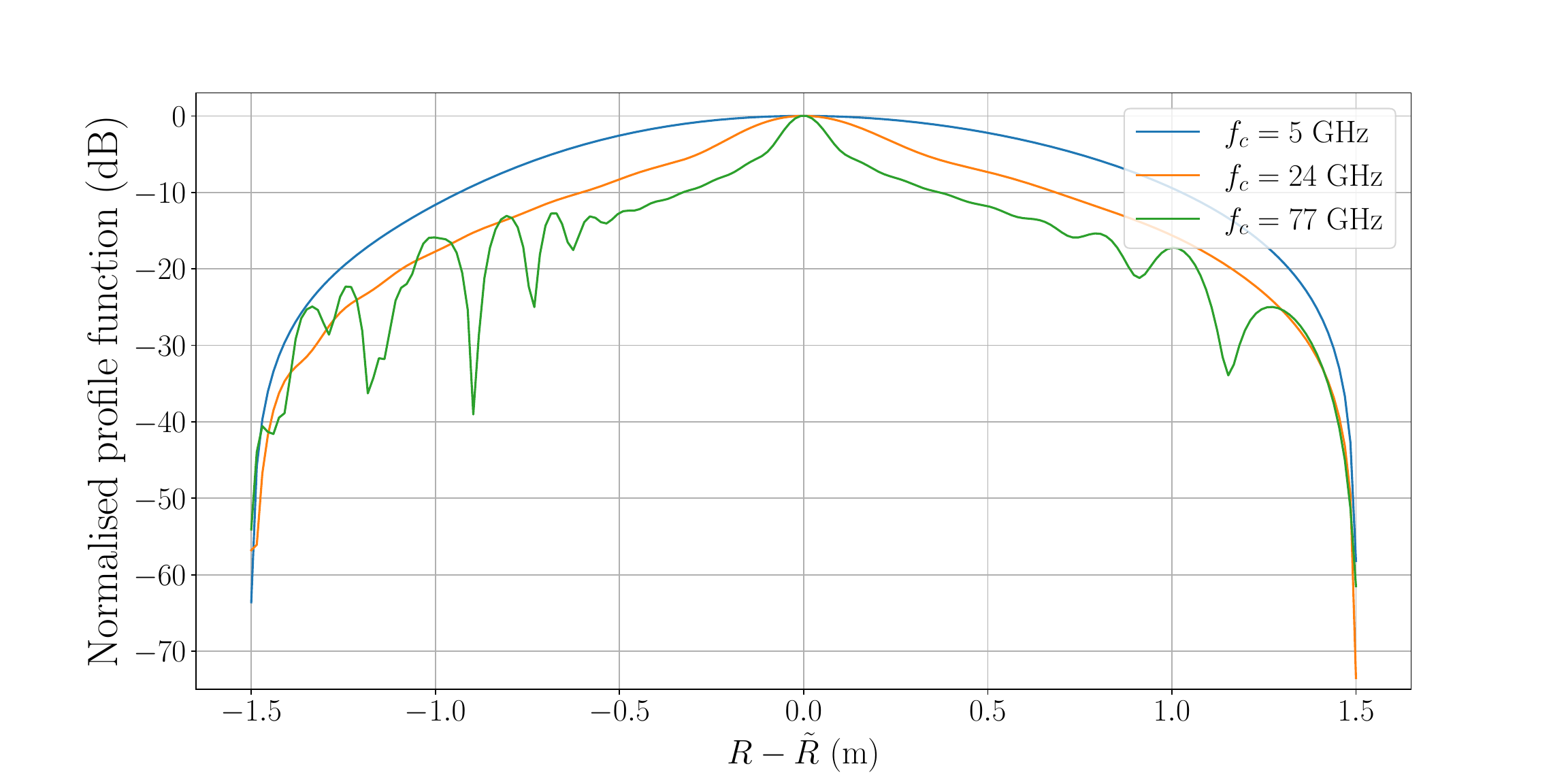}
            \vspace{0.01pt}
        \end{subfigure} &
        \begin{subfigure}[t]{0.3\textwidth}
            \centering
            \includegraphics[width=\linewidth, trim=60 0 100 60, clip]{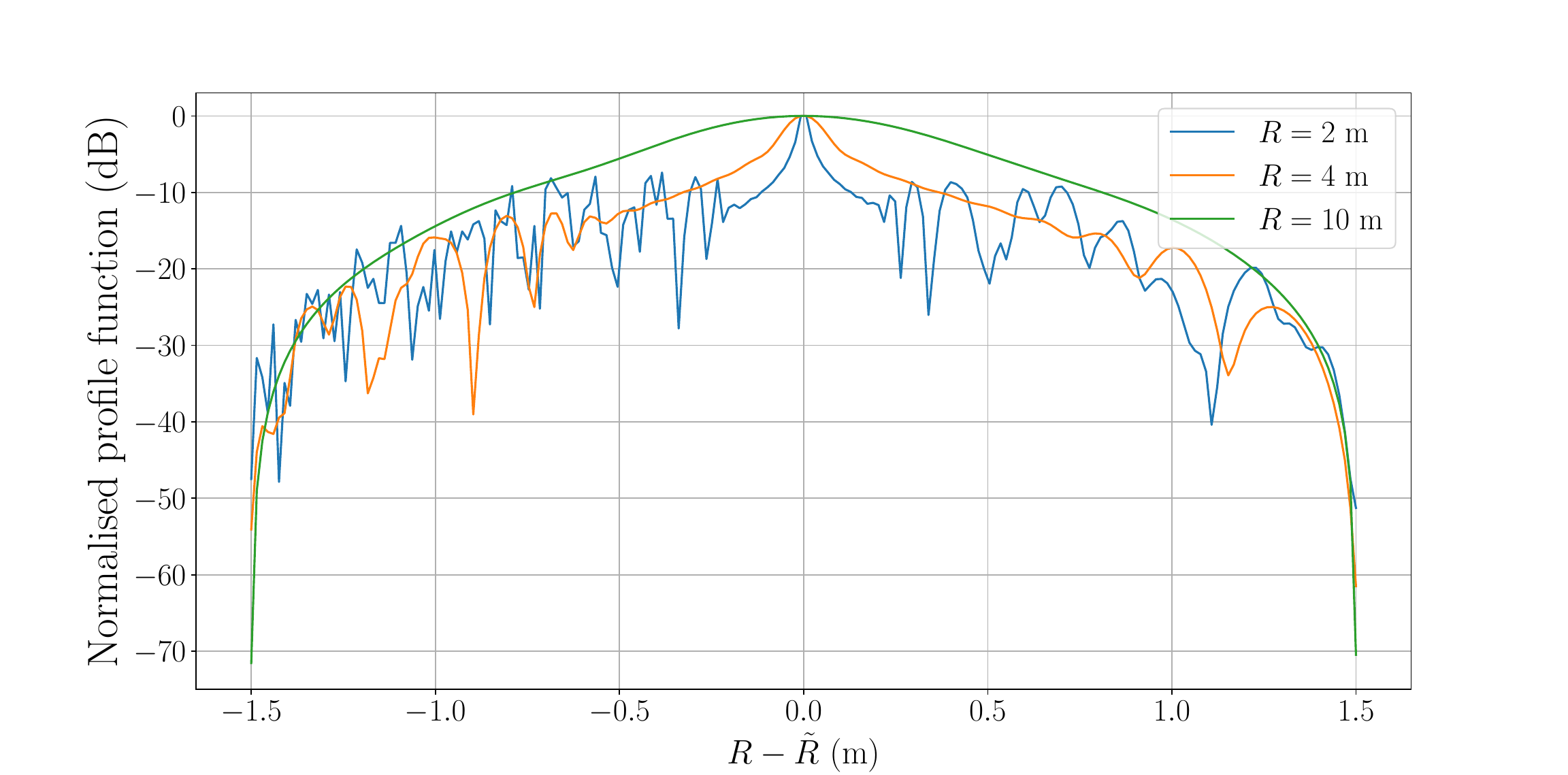}
            \vspace{0.01pt}
        \end{subfigure} \\
        \begin{subfigure}[t]{0.3\textwidth}
            \centering
            \includegraphics[width=\linewidth, trim=60 0 100 60, clip]{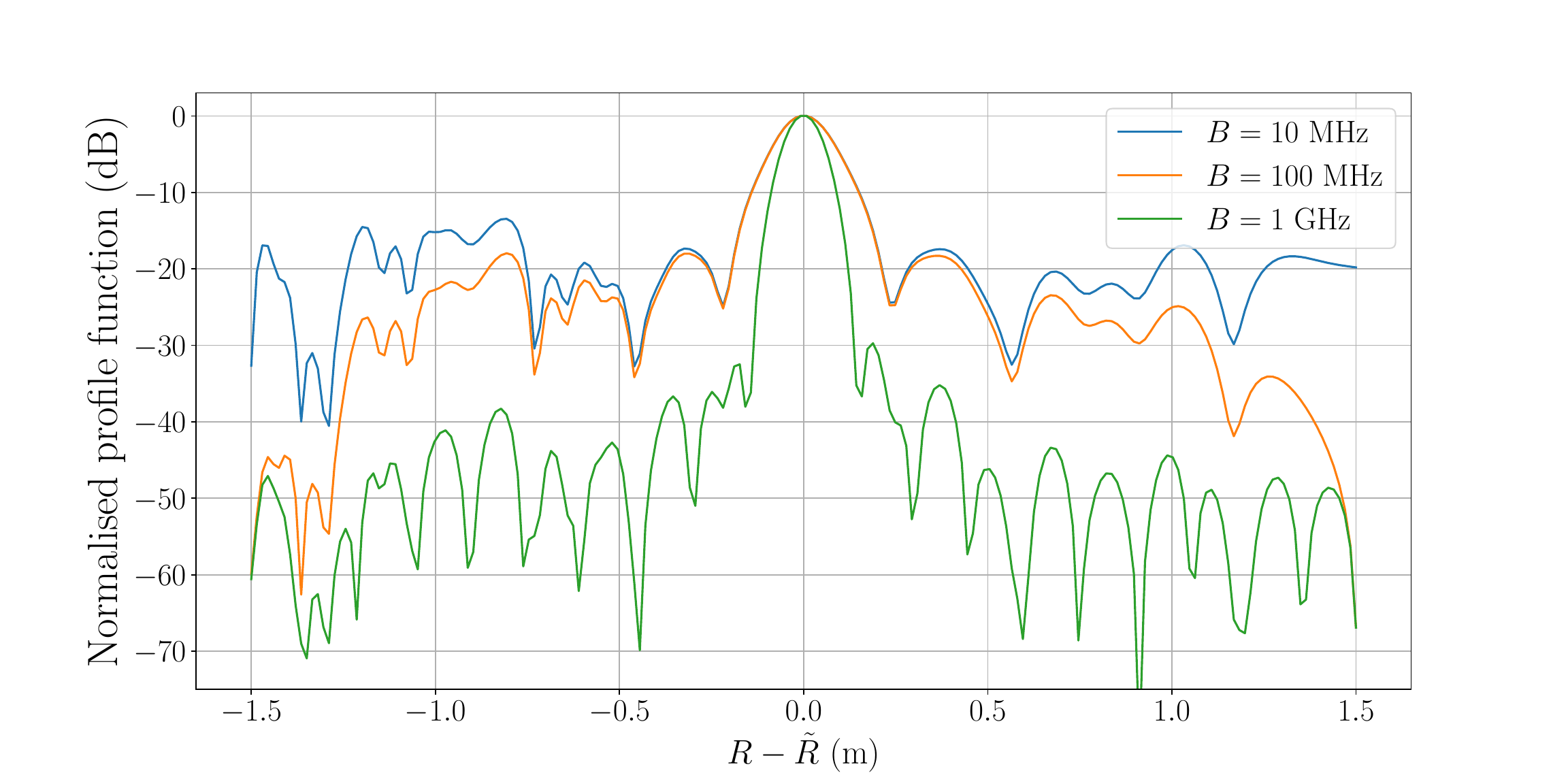}
        \end{subfigure} &
        \begin{subfigure}[t]{0.3\textwidth}
            \centering
            \includegraphics[width=\linewidth, trim=60 0 100 60, clip]{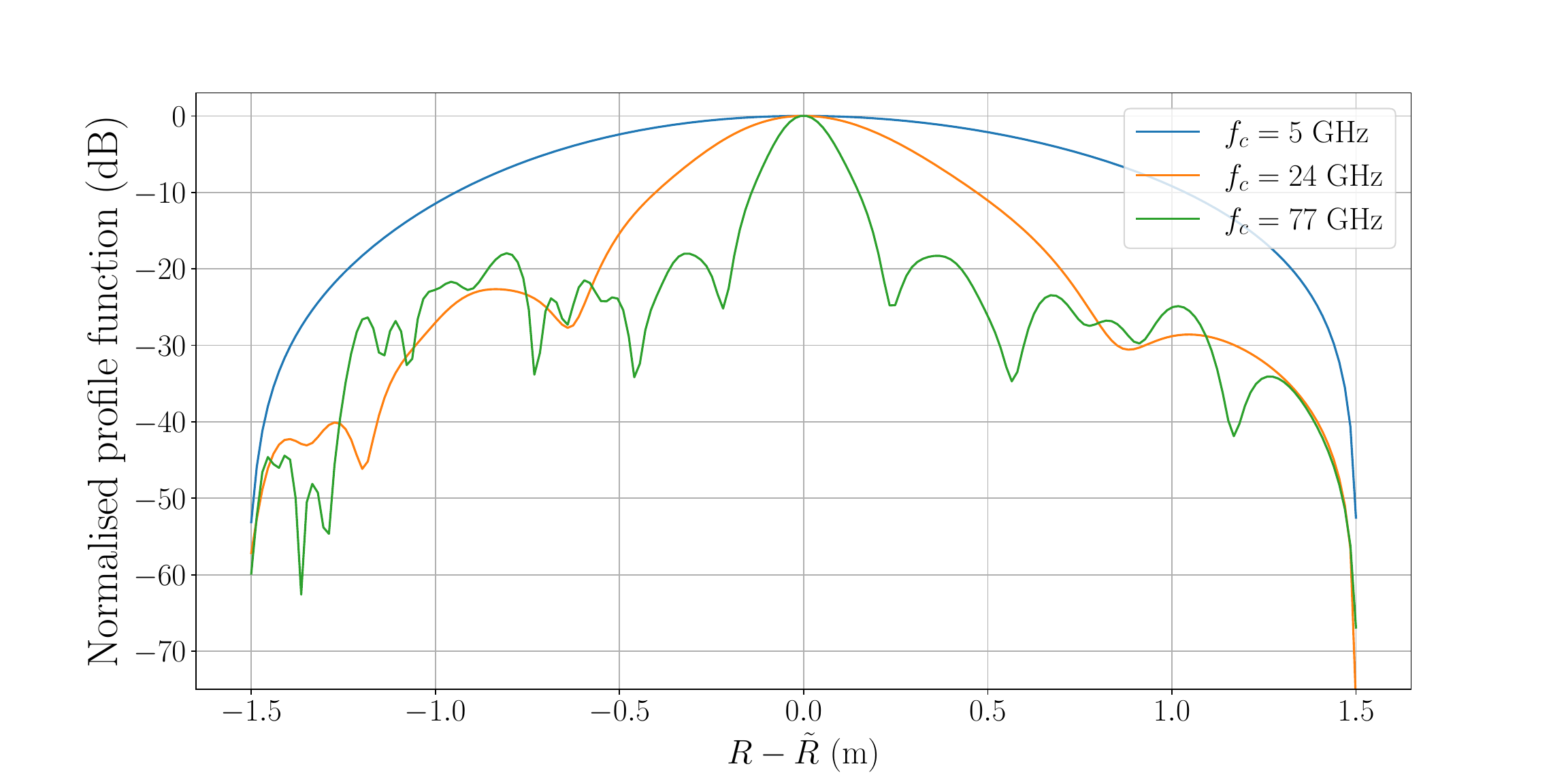}
        \end{subfigure} &
        \begin{subfigure}[t]{0.3\textwidth}
            \centering
            \includegraphics[width=\linewidth,trim=60 0 100 60, clip]{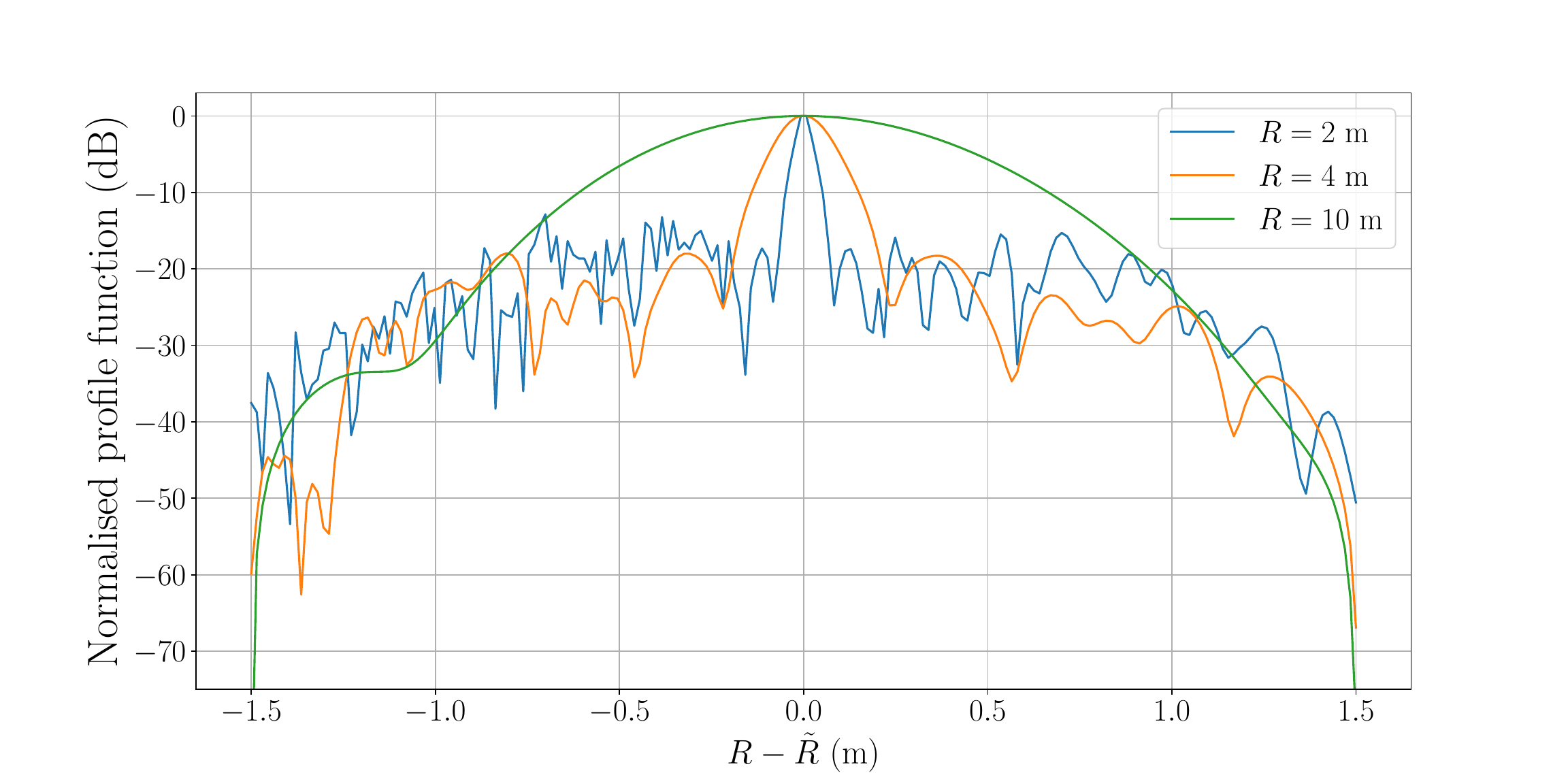}
        \end{subfigure} \\

    \end{tabular}

    \begin{picture}(0,0)
        \put(-260,120){\rotatebox{90}{Flat Plate}}
        \put(-260,32){\rotatebox{90}{Sphere}}
    \end{picture}

    \caption{Impact of the bandwidth $B$, carrier frequency $f_c$ and range $R$ on the range profile functions of the two canonical targets. }
    \label{fig:impact_parameters}
\end{figure*}

As shown in Figure~\ref{fig:impact_parameters}, similar results to those obtained in \cite{moulin_near-field_2024} can be observed with flat and curved targets. Indeed, the closer the target, the narrower the main lobe. Additionally, higher bandwidths and carrier frequencies improve the resolution, here defined as the width of the main lobe at -3 dB. Nonetheless, a significant decrease of the side lobes level is observed with curved geometries. It becomes less likely to confuse the true maximum with a noise-enhanced secondary maximum, causing the flat geometry to be more prone to precision errors. At low Signal-to-Noise (SNR) ratio,  with the flat geometry, owing to the null elevation and azimuth angles and the fact that all specular points lie on the target, many antenna pairs receive identical signals. Compared to curved geometries, this results in redundant information, governed by the relative positions of the transmit and receive antennas.

\section{Extended vs Point Target Models}

\label{sec:stationary_points_analysis}

This section investigates the effect of model mismatch between the extended and point target models on range estimation. 

\subsection{Profile function}

\begin{figure}
    \centering
    \includegraphics[width=0.5\textwidth]{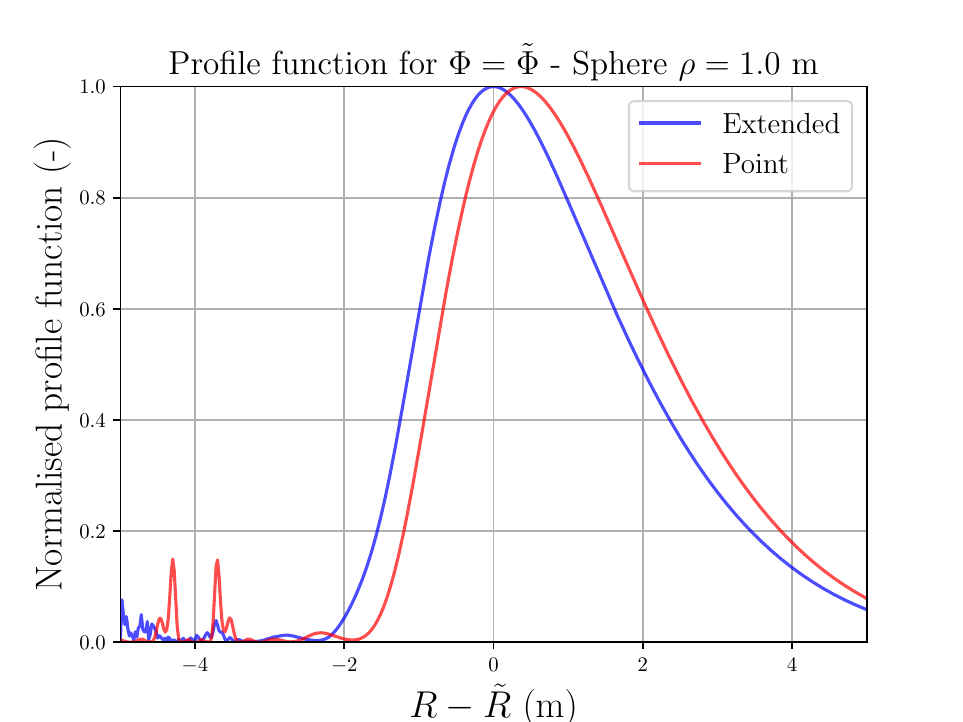}
    \caption{Profile functions of the extended and point target models for a spherical target.}
    \label{fig:profile_function_mismatch_example}
\end{figure}

Let us first compare the normalised profile functions obtained with the point and extended target models for a spherical target of radius $\rho=1 \, \text{m}$, as shown in Figure~\ref{fig:profile_function_mismatch_example}.
The simulation parameters are given in Table~\ref{tab:parameters}, except for the carrier frequency and bandwidth, which are here set to $f_c=3.5 \, \text{GHz}$ and $B=18 \, \text{MHz}$ respectively, to better highlight the model mismatch. 
In the point target model, the target is represented as a single point at the origin with coordinates $(0, 0, 0)$. For the extended target model, the target is centred at $(\rho, 0, 0)$ in the curved case and at $(0, 0, 0)$ in the flat case. 

It can be seen that the maximum of the profile function for the point target model does not correspond to the correct range, whereas this is the case for the extended NF EM model. 
Therefore, the point target model does not estimate the correct distance between the centre of the antenna array and the target, even in the absence of noise. This is because treating the target as a point causes the model to consider the distances between antennas passing through the centre point instead of the specular point for each pair of antennas. 

\subsection{Range Model Mismatch}

\begin{figure}
    \centering
    \includegraphics[width=0.5\textwidth]{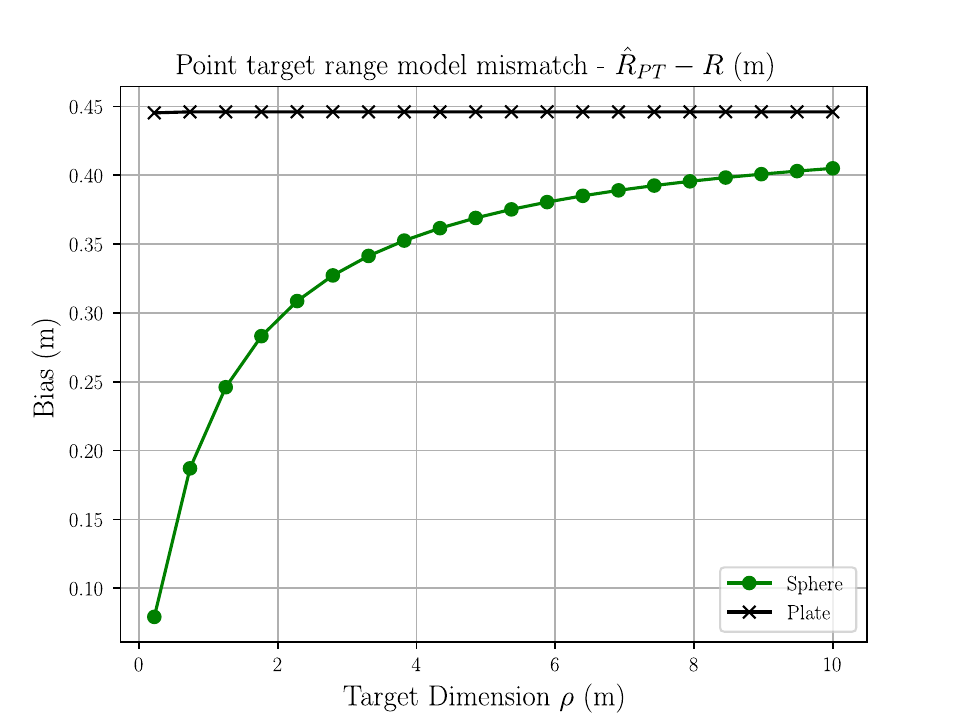}
    \caption{Point target model mismatch for increasing dimensions of plate and spherical targets at range $R=1 \, \text{m}$ and antenna spacing $\Delta=\lambda/2\approx 0.086 \, \text{m}$.}
    \label{fig:mismatch_analysis_rho_plate_sphere}
\end{figure}

Then, let us compare the range obtained with a sphere and a flat plate while increasing the target dimension. The results are shown in Figure~\ref{fig:mismatch_analysis_rho_plate_sphere}. In this case, $\rho$ denotes the radius for the sphere or half the length for the plate. For example, at a distance  $R=1 \, \text{m}$ and antenna spacing $\Delta=\lambda/2\approx 0.086 \, \text{m}$, when all the stationary points are on the plate -- that is, when the plate length is equal to the array size -- the distance estimation error for the flat target stabilises at $0.446 \, \text{m}$. This error remains constant as the plate size increases because the stationary points do not shift with larger target sizes. 
In contrast, the distance estimation error for a sphere increases with its radius, reflecting its increasing similarity to a plate. However, the error for the sphere remains below that of the plate and asymptotically converges to it as the radius of the sphere increases. 

\subsection{Stationary Point Analysis}

The distribution of the specular points for the three canonical targets, i.e. a flat rectangular target, a sphere, and a cylinder, is analysed in this section. A $10 \times 10$ antenna array with an antenna spacing of $0.1 \, \text{m}$ to cover the complete target surface, oriented parallel to the $y-z$ plane, is positioned at a distance $R = 4 \, \text{m}$ from the target. The target has dimensions $\ L = D_y = D_z = 1 \, \text{m}$ and its curvature radius is equal to $\rho = 0.707 \, \text{m}$. Note that the dimensions of the target does not affect the extended model under the SPA.
\begin{figure*}
    \centering
    \includegraphics[width=\textwidth]{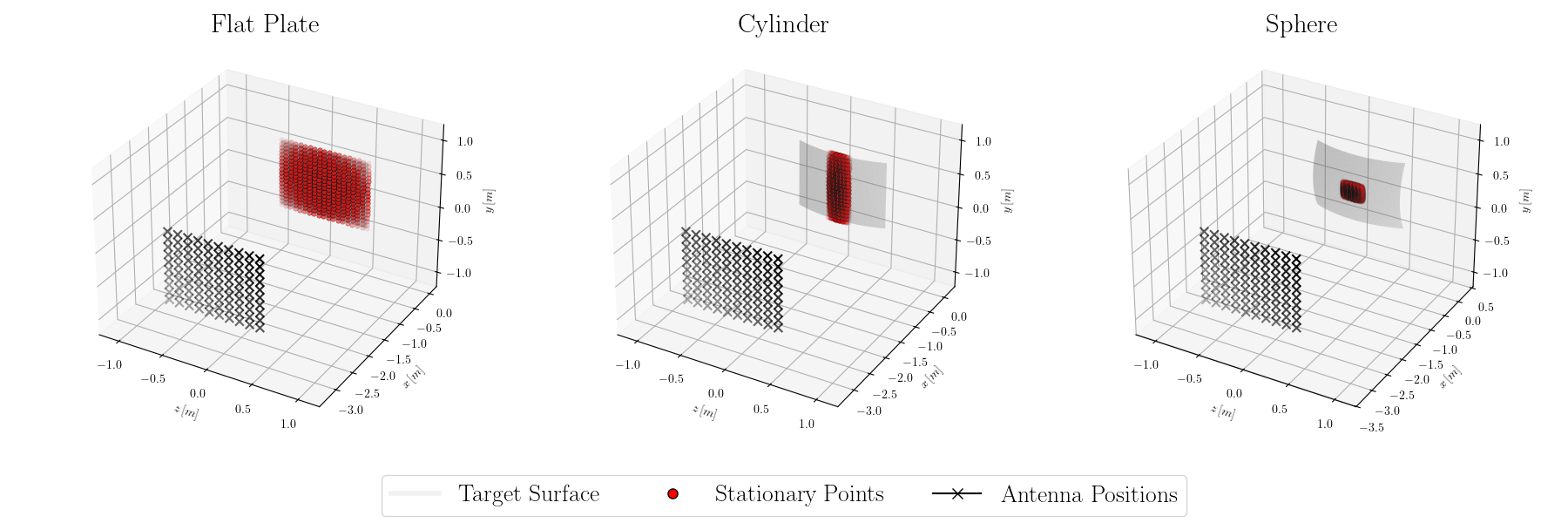}
    \caption{Analysis of Stationary Points for the three canonical targets. The antenna positions and specular points locations are respectively depicted by black crosses and red dots.}
    \label{fig:stationary_points_analysis}
\end{figure*}

Figure \ref{fig:stationary_points_analysis} shows the spatial distribution of stationary points between each pair of antennas for the three types of targets.
The stationary points are relatively evenly distributed over the surface of the flat rectangular target. For the cylindrical target, despite its \(1 \, \text{m}\) diameter, the stationary points show a higher concentration near the centre along the \(z-\)axis (axis of curvature), while maintaining a more uniform distribution along the \(y-\)axis. In the case of the spherical target, the stationary points are more densely clustered near the point on the sphere closest to the antenna array. With curvature, the stationary points tend to be more concentrated around a single line (resp. point) in the cylinder (resp. sphere) case. This is a consequence of the curvature of the target, which causes the phase to vary more rapidly, resulting in more localised stationary points. 

Consequently, the point target assumption seems to remain reasonable for curved objects, as the stationary points are more concentrated around a single point for curved targets. Yet, this point is not located at the centre, but on the surface of the target! 

\subsection{Summary}

These results show the emergence of a range bias in the profile function of the point target model for extended targets. This discrepancy results from the assumption for the point target model that all antenna pairs interact with a single point, rather than accounting for the distributed nature of the target. 
The observed variation in model mismatch with the target shape can be intuitively explained by the curvature-induced concentration of stationary  points.

\label{sec:extended_vs_PT}

\section{Model Mismatch Characterisation}
\label{sec:model_mismatch_characterisation}

This section provides a detailed characterisation of the range estimation bias for the point target model as a function of system parameters.

\subsection{Considered Estimator} 

In order to evaluate the estimation bias with the point target model, an alternative range estimator is considered. Compared to the initial estimator proposed in \cite{moulin_near-field_2024}, the alternative estimator enables to develop an analytical expression for the bias to be derived, eliminating the need for an exhaustive search of the maximum of the profile function. 

This estimator is based on the least square error minimization of the total travelled distances between each pair of antennas and can be formulated as 
\begin{equation}
    \hat{R} = \argmin_{\tilde{R}} \sum_{l'=0}^{N-1} \sum_{l=0}^{N-1} ( r_{ls} + r_{l's} - (\tilde{r}_{ls;\tilde{R}} + \tilde{r}_{l's;\tilde{R}}) )^2, 
\label{eq:ls_minimization}
\end{equation}
where \( r_{ls} \) and \( r_{l's} \) represent the actual distances to the stationary point for the ($l'$;$l$) antenna pair, while \( \tilde{r}_{ls;\tilde{R}} \) and \( \tilde{r}_{l's;\tilde{R}} \) denote their modelled counterparts, which depend on the test range $\tilde{R}$ and the selected model (extended or point target). In the case of a point target model, the candidate distances no longer depend on the stationary points but on the point chosen to represent the entire target, and could be rewritten \( \tilde{r}_{l;\tilde{R}} \) and~\( \tilde{r}_{l';\tilde{R}} \). 

In practice, the estimation is performed in two steps. First, the total travelled distance $ \hat{r}_{ll'} = \hat{r}_{l} + \hat{r}_{l'}$ is estimated for each antenna pair using the received signals  
\begin{equation}
u_{l'l}(t) = A_{l'l} e^{j\psi(\textbf{x}_{s,l'l})} s_{l}\left(t- \frac{r_{ls} + r_{l's}}{c}\right),
\end{equation}  
where \( A_{ll'} \) is the amplitude factor incorporating the channel effects introduced in Sections~\ref{sec:system_model} and \ref{sec:SPA}.  
Then, the real distances are replaced by the estimated distance in \eqref{eq:ls_minimization}.  
The problem, as formulated in \eqref{eq:ls_minimization} with the actual travelled distances, therefore represents an idealised scenario %
and corresponds to a "genie"-aided version.
Although this approximation does not hold in practical scenarios, it is still considered here as it allows an analytical expression of the bias to be derived, which facilitates its interpretation and provides bias values that are independent of the specific estimator that would be used to estimate the total travelled distance.

It can be shown that for the same set of parameters, this ideal estimator gives lower biases for the point target model. This is due to the prior removal of phase-related ambiguities and uncertainties by the use of actual inter-antenna distances rather than estimated ones, thus providing more accurate range estimates compared to conventional ML approaches. As a result, the model mismatch obtained with this hypothetical estimator serves as a lowest bound, meaning that no lower model mismatch can be expected when using the ML estimator described in \cite{moulin_near-field_2024} and used above.  

\subsection{Numerical Analyses}

This subsection presents various numerical analyses of the range mismatch, based on the range estimator introduced in the previous subsection.

\subsubsection{Linear Antenna Array}

First, the analysis of the range mismatch as a function of the system parameters is performed for a linear antenna array. 

\begin{figure*}[ht]
    \centering
    \begin{subfigure}[t]{0.32\textwidth}
        \centering
        \includegraphics[width=1.2\textwidth]{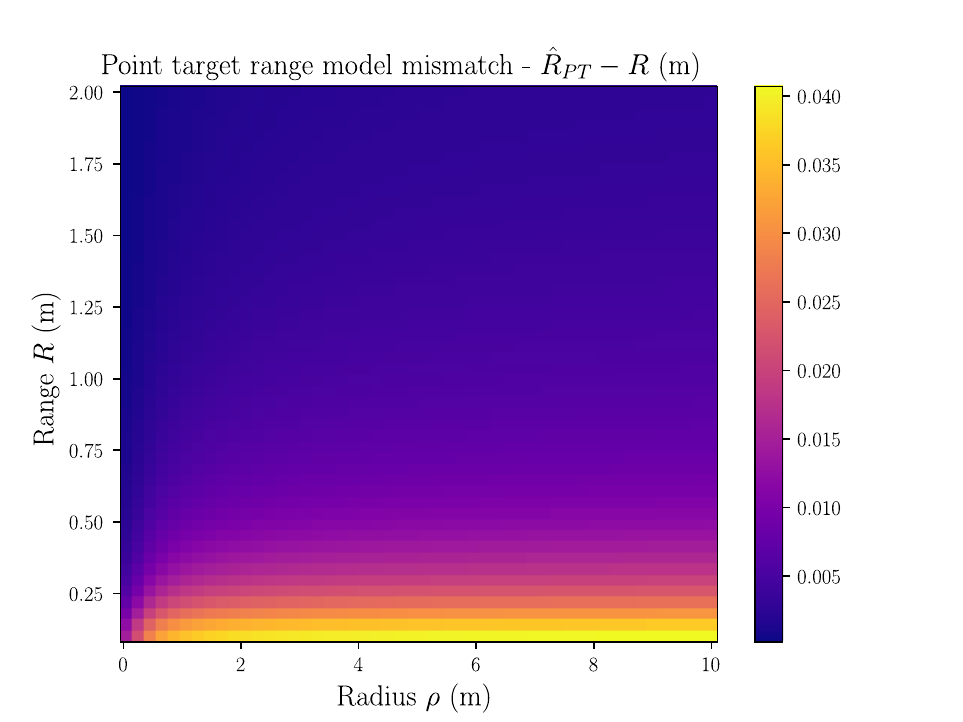}
        \caption{Range - radius, antenna spacing $\Delta=\lambda/2\approx 0.086 \, \text{m}$. \\}
        \label{fig:sub1}
    \end{subfigure}
    \hfill
    \begin{subfigure}[t]{0.32\textwidth}
        \centering
        \includegraphics[width=1.2\textwidth]{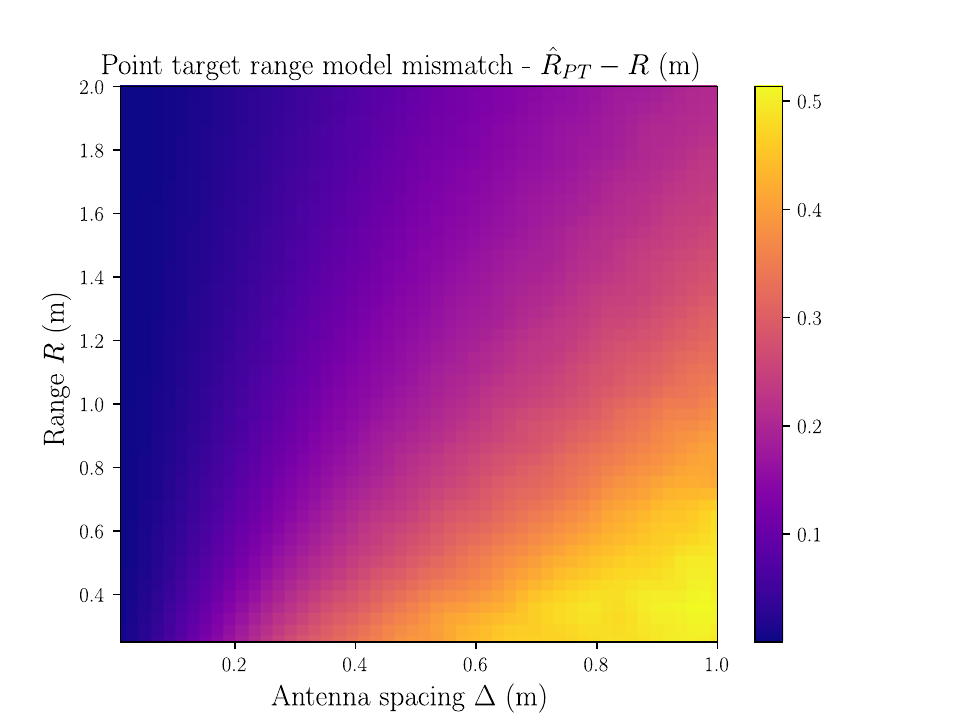}
        \caption{Range - antenna spacing, radius $\rho = 1 \, \text{m}$. \\}
        \label{fig:sub2}
    \end{subfigure}
    \hfill
    \begin{subfigure}[t]{0.32\textwidth}
        \centering
        \includegraphics[width=1.2\textwidth]{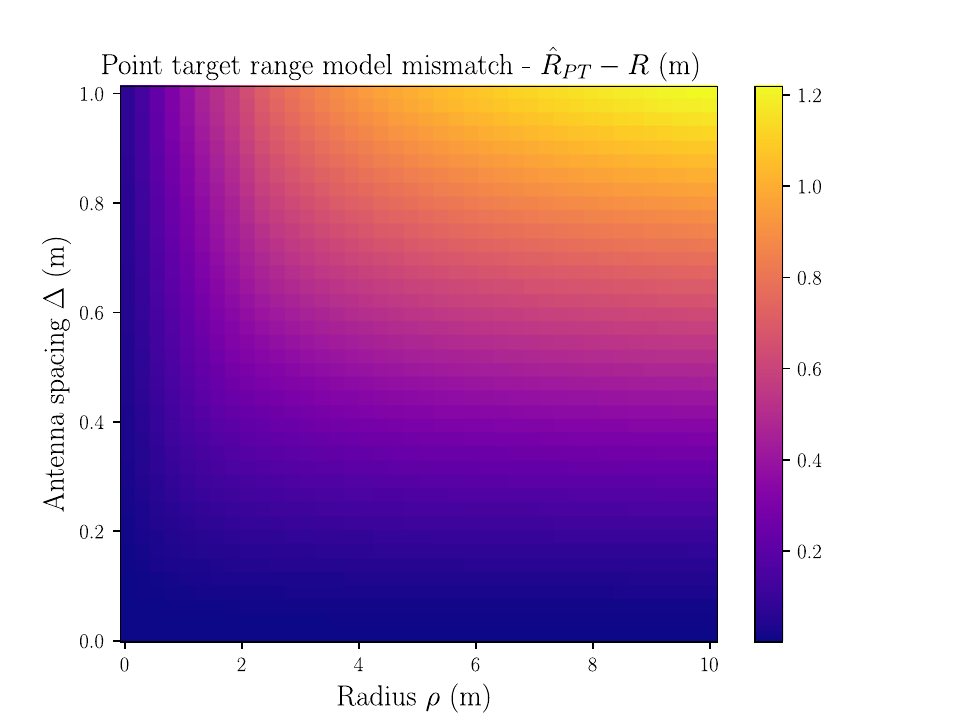}
        \caption{Antenna spacing - radius, range $R=1 \, \text{m}$. \\}
        \label{fig:sub2}
    \end{subfigure}
    \caption{Range bias analysis for the point target model and the spherical target as a function of the sphere radius \( \rho \), the range \( R \), and the antenna spacing \( \Delta \) using the least squares error minimization estimator.}
    \label{fig:pt_mismatch_analysis_parameters}
\end{figure*}

For a linear array oriented along the $z$-axis and positioned at $y = 0$, the distance between an antenna at position $\mathbf{x}_l$ and range $R$, and a point $P$ at position $\mathbf{x}_P$ can be computed using the law of cosines, followed by a Taylor series expansion, as 
\begin{equation}
    \| \mathbf{x}_p - \mathbf{x}_l \| \approx \tilde{r}_{l;R} + \frac{r_P^2}{2\tilde{r}_{l;R}} - r_P  \cos(\phi_l - \phi_P), 
\end{equation}
where $\phi_l$ and $\phi_P$ represent the azimuth angles, and $\tilde{r}_{l;R}$ and $r_P$ denote the distances from the antenna and point $P$ to the origin of the coordinate system, respectively.
Under the per-antenna far-field (FF) assumption, which is valid when $R \gg r_P$, and assuming that the target representative point is located at $\theta_P={\pi}/{2}$, the estimation of the range for a single point comes to estimating the distance $r_P$, i.e. $\hat{r}_P=\hat{R}$.  Consequently, the range mismatch for all antenna pairs, based on the range estimation defined in \eqref{eq:ls_minimization}, can be expressed as
\begin{equation}
        \hat{R} - R =
        \frac{
            \sum_{l=0}^{N-1} \sum_{l'=0}^{N-1} 
            \left( \left(\tilde{r}_{l;R}+\tilde{r}_{l';R}\right) 
            - \left(r_{ls}+r_{l's}\right) \right) 
            \left(\sin \theta_l + \sin \theta_{l'}\right)
        }{
            \sum_{l'=0}^{N-1} \sum_{l=0}^{N-1} 
            (\sin \theta_l + \sin \theta_{l'})^2
        }. 
\label{eq:mismatch_analytical}
\end{equation}
The mismatch for the point model comes from considering the distances $\tilde{r}_{l;R}$ and $\tilde{r}_{l';R}$ to the single reference point representing the target for all pairs of antennas,  rather than using the stationary points. The bias obtained with the point target model can be seen as a weighted sum of the discrepancies between the actual propagation distances and those to the reference point. It can also be noted that the extended target model does not introduce any bias as it accurately accounts for the distances to the stationary points. 

Figure~\ref{fig:pt_mismatch_analysis_parameters} further investigates the range estimation bias due to model mismatch for a spherical target with this idealistic estimator, leveraging \eqref{eq:mismatch_analytical}. The range bias evolution is shown as a function of the range $R$, sphere radius $\rho$ and antenna spacing~$\Delta$. The analysis is based on a linear array of 13 antennas. Note that the colour scales are not identical across the sub-figures. The impact of the system parameters can be described as follows:

\begin{itemize}

\item \textit{Range $R$:} As the range increases, the mismatch decreases because the spherical target appears more like a single point from the perspective of the antenna array.

\item \textit{Antenna Spacing $\Delta$:} Greater biases are obtained as the distance between the antennas increases, corresponding to a larger array size. This effect occurs because, as the array expands, the discrepancy between the actual distances to the stationary points and those to the representative target point becomes more pronounced since the stationary points move further away from the assumed reference point.

\item \textit{Radius $\rho$:} As the sphere size decreases, this distance difference diminishes because the stationary points are closer to this abstracting point. However, if the radius of curvature of the sphere increases, the estimation error due to mismatch also increases. In this case, a flatter geometry from the perspective of the antenna array results in an increased mismatch. 

\end{itemize}

In the case of a large antenna array, the expected mismatches can thus become significant, particularly if the target is large and at close range. On the contrary, for a more distant target, with a smaller antenna array or reduced size target, abstracting the target to a single point seems more reasonable.

\subsubsection{Distributed Antennas}

The previous subsection analysed the range mismatch for a given antenna array. However, the developed electromagnetic target model is also applicable to distributed antenna configurations. Therefore, this subsection extends the mismatch analysis to this scenario. 

For this analysis, three 13-element antenna arrays aligned along the \( z \)-axis and positioned at \( y = 0 \) are considered. Within each sub-array, the antennas are spaced at \( \lambda/2 \approx 0.086 \, \text{m}\), corresponding to a carrier frequency of \( f_c = 3.5\,\mathrm{GHz} \).

\begin{figure}
    \centering
    \includegraphics[width=0.5\textwidth]{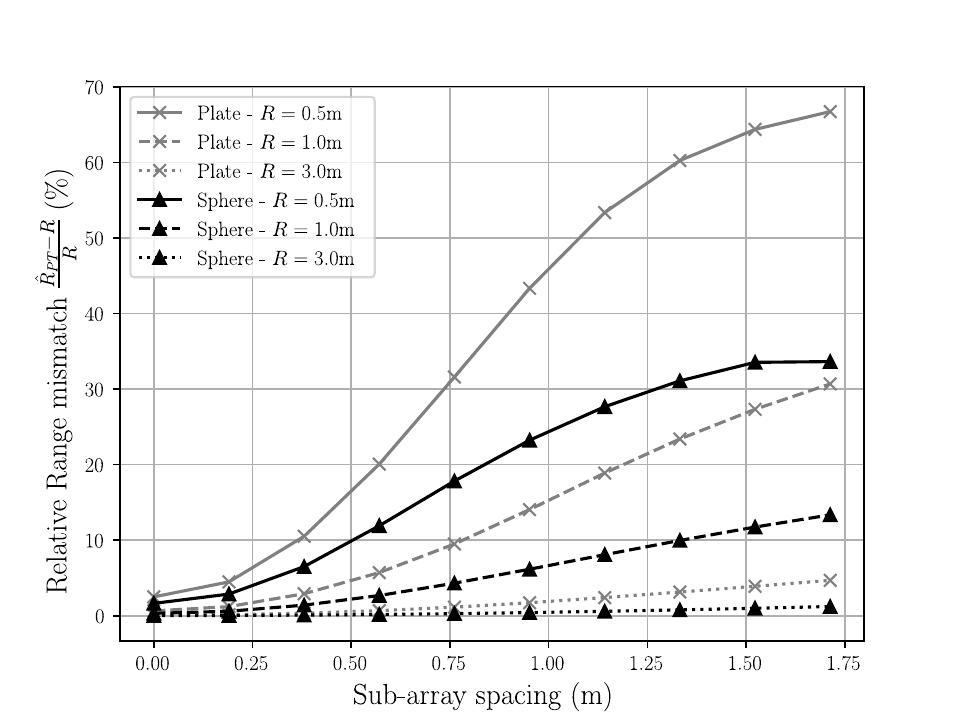}
    \caption{Relative range mismatch in a distributed antenna array configuration of three 13-antenna arrays as a function of sub-array spacing for a spherical target and rectangular plate.}
    \label{fig:distributed_mismatch}
\end{figure}

Figure~\ref{fig:distributed_mismatch} illustrates the relative range mismatch as a function of the sub-array spacing. The central sub-array is positioned at \( z = 0 \), facing the target, while the other two sub-arrays are positioned symmetrically along the \( z \) axis -- one shifted in the positive \( z \) direction and the other shifted in the negative direction by a specified sub-array spacing measured from center to center. Two types of targets are considered at different ranges $R$: a spherical target and a flat rectangular plate, each with a radius or half-length of \( \rho = 1\,\text{m} \).

Similar to the single array case, the distributed configuration shows a higher range mismatch for the flat target compared to the curved target for the same set of parameters. 
As the system becomes more distributed (i.e. as the sub-array spacing increases), the relative mismatch becomes more significant. This behaviour is attributed to the fact that the stationary points become more dispersed, deviating from the single point approximation. An increase in range leads to a reduction in relative mismatch levels, consistent with the observations and explanations in the previous subsection but here for a distributed scenario.

Therefore, depending on the scenario, the model mismatch can vary significantly, from negligible in cases of curved targets and small arrays at greater distances, to detrimental in scenarios with flatter, closer targets and more distributed antenna configurations. In this context, analyses similar to Figures~\ref{fig:pt_mismatch_analysis_parameters} and \ref{fig:distributed_mismatch} can be used to estimate the minimum expected model mismatch or relative error for a given scenario.

\subsubsection{Equipotential Curves}

Equipotential curves, i.e. curves for which a given set of system parameters gives the same range mismatch, can be obtained by equating \eqref{eq:mismatch_analytical} to a specified bias value \( \alpha \). Besides, if a certain range mismatch maximum threshold \( \alpha \) is defined as acceptable, these curves delineate the limits of the operating region within which the point-target approximation remains valid.

For example, for a given antenna array, assuming that the antenna spacing is small compared to the range, i.e. $R >> \Delta$, the equipotential curve equation for a given range bias $\alpha$ in the case of a flat plate target and a single antenna array is given by
\begin{equation}
    R = \frac{
        \sum_{l'=0}^{N-1} \sum_{l=0}^{N-1} 
        \left( \frac{(l - \frac{N - 1}{2})^2 + (l' - \frac{N - 1}{2})^2}{2} - \frac{(l - l')^2}{2} \right)
    }{
        2 N^2 \alpha 
    }  \Delta^2,  
\label{eq:potential_curve_plate}
\end{equation}
where all stationary points are assumed to be located on the plate. 

\begin{figure}
    \centering
    \includegraphics[width=0.5\textwidth]{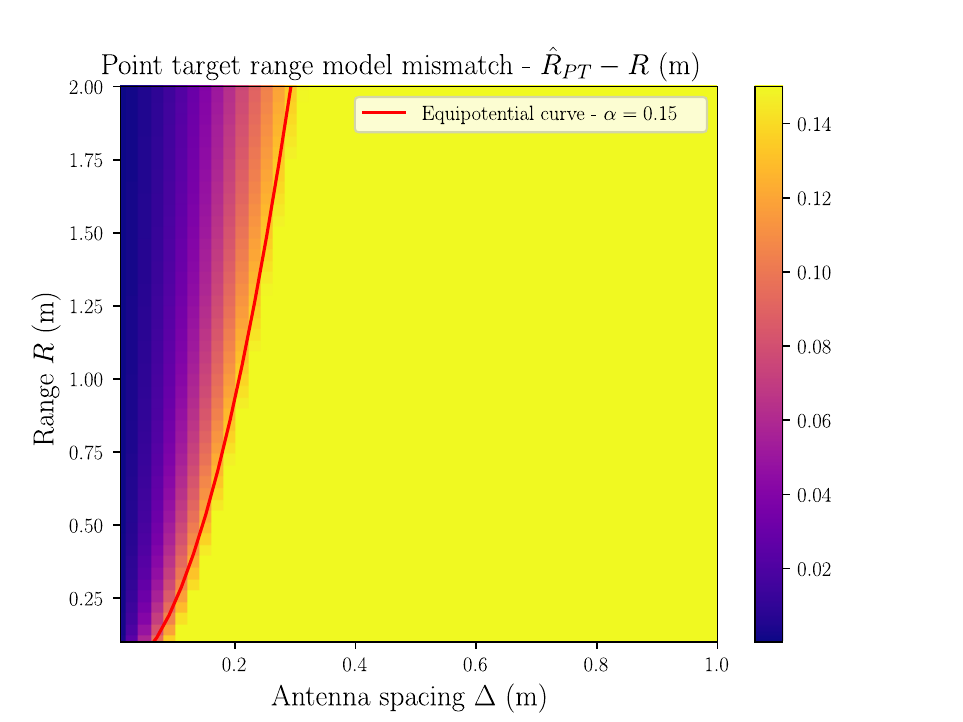}
    \caption{Equipotential curve for the range mismatch analysis as a function of the range $R$ and antenna spacing$\Delta$ for a flat plate.}
    \label{fig:potential curve}
\end{figure}

An illustration of such curve is given in red in Figure~\ref{fig:potential curve} and illustrates the quadratic relation between the range and antenna spacing at constant bias for $\alpha=0.15 \, \text{m}$. It can be seen that the analytical curve derived from \eqref{eq:potential_curve_plate} fits well with the thresholded mismatch colour map obtained with \eqref{eq:mismatch_analytical}. In this way, \eqref{eq:potential_curve_plate} can be used to determine the range-antenna spacing parameter pairs $(R;\Delta)$ that give rise to the same bias in the range estimate with a point target model of a rectangular flat plate. Moreover, if the range mismatch threshold \( \alpha \) is predefined as permissible, this curve also delineates the operational zone within which the point target approximation remains valid. In Figure~\ref{fig:potential curve}, this corresponds to the (non-yellow) region at the left of the red curve. 

For other target geometries, such as the sphere or cylinder, obtaining an analytical expression for the potential curves is more complicated. In such cases, analyses based on the colour maps shown in Figure \ref{fig:pt_mismatch_analysis_parameters} can be performed by looking at the zones where the range mismatch is equal to a given value $\alpha$. 

\subsubsection{Summary}

To summarise, Figures~\ref{fig:mismatch_analysis_rho_plate_sphere}, \ref{fig:stationary_points_analysis} and \ref{fig:pt_mismatch_analysis_parameters}  show that the point target approximation performs well when the dimensions of the target are relatively small w.r.t. the array size, and it exhibits curvature. In this context, the flat target scenario represents a worst-case condition for point-target approximation, explaining the limitations raised in \cite{thiran_performance_2024}.
This phenomenon also explains the accurate localisation obtained experimentally by \cite{sakhnini_near-field_2022} and \cite{sakhnini_experimental_2022} when considering a small cylinder as single point. However, for larger target or antenna array dimensions, specular points have a greater tendency to deviate from the single point case and larger model mismatch estimation errors can occur, as illustrated in Figures~\ref{fig:pt_mismatch_analysis_parameters} and \ref{fig:distributed_mismatch}. To determine to which extent the point target assumption remains valid, equipotential curves similar to \eqref{eq:potential_curve_plate} and Figure~\ref{fig:potential curve}, can be~used.

\section{Conclusion}
This paper provides a comprehensive analysis and validation of an EM-based signal model for an arbitrary target in the NF region of an arbitrary network of antennas, based on the SPA method. Further developments are made to provide mathematical solutions for three canonical targets, i.e. flat rectangular plate, sphere, and cylinder. 
First, the profile function analysis of a range ML estimator showed high accuracy, with target curvature effectively reducing side lobes.
Furthermore, a model mismatch analysis with the point target model showed an increasing estimation error with increasing array size and radius of curvature, and a stationary point analysis revealed a concentration of stationary points at the centre of curved targets. 
These results highlight the relevancy of point target models, and explains the accurate localisation observed in other near-field sensing papers, when small curved objects are approximated as point targets \cite{sakhnini_experimental_2022, sakhnini_near-field_2022}. Nonetheless, as soon as the target dimensions become relatively large compared to the antenna array size, this approximation is no longer valid and limitations must be taken into account. 
The operational regions where the point target assumption holds or fails for a given mismatch value have been derived.  

Future work will explore estimation of the target shape and sensing of multiple targets, and the study of efficient algorithms for 3D localisation.

\bibliographystyle{ieeetr}
\bibliography{biblio}

\appendices

\end{document}